# Dynamical Process of Liner Implosion in the Electromagnetic Flux Compression for Ultra-high Magnetic Fields

D Nakamura, H Sawabe, Y H Matsuda and S Takeyama

Institute for Solid State Physics, University of Tokyo, 5-1-5, Kashiwanoha, Kashiwa, Chiba 277-8581, Japan

E-mail: dnakamura@issp.u-tokyo.ac.jp

**Abstract**
The spatial distribution of magnetic fields that are generated by the electromagnetic flux compression technique is investigated, with emphasis on the dynamical processes of an imploding liner. By comparing with the results of computer simulations, we found that the non-uniform implosion of a liner is important in order to explain the magnetic field's distribution during the liner's implosion. In addition, our results suggest that the initial inwards compressing spool-like motion of the liner subsequently turns out to be outwards stretching barrel-like motion along the magnetic field axis.
PACS numbers: 07.55.Db, 07.05.Tp

## 1. Introduction

Generation of ultra-high magnetic fields with intensities above 100 T necessitates the use of specialized techniques, since there is an unavoidable destructive effect due to the Maxwell stress ($> 40,000$ kg/cm$^2$) that exceeds the yield stress acting on a magnet coil [1-3]. Magnetic fields with intensities of ~1000 T can be generated by using the magnetic flux compression techniques, in which the magnetic field is compressed within a few microseconds before the destruction of the magnet. As an example, in such ultra-high magnetic fields the magnetic length (or twice the cyclotron radius) approaches the nanoscale range. The Landau-level splitting of a carrier exceeds the thermal fluctuation energy, and, as a result, diverse quantum phenomena are observable at temperatures as high as room temperature.

Several issues need to be addressed before reliable measurements for the solid-state physics can be successfully performed in an environment of destructive magnets. In the flux compression technique, the initial magnetic field that is generated by a seed field coil is compressed by an imploding metallic cylinder that is called the 'liner'. Stability of the imploding liner's centre position and high reproducibility are required, because by the time reaching the peak field, the imploding liner shrinks to a structure that has an inner radius of a few millimetres. The explosive-driven flux compression method [4,5] that makes use of chemical explosives to accelerate the implosive liner, renders a significant hurdle on the way of the application of this method to the measurements in a solid-state physics experiment, due to the violent destruction and resulting inferior reproducibility that are associated with this method.

On the other hand, the electromagnetic flux compression (EMFC) method [6] utilizes the inductive electromagnetic force to accelerate a liner, which arises from the electrical current injected into the primary coil set outside the liner. This technique is more suitable for application to solid-state physics experiments due to the more precise controllability as compared with the chemical explosive-driven method. The EMFC technique was developed in Institute for Solid State Physics (ISSP) over 30 years. Especially, the design of the primary coil is essential for effective induction of the current of up to several mega-amperes into the liner, in the range of microseconds. In a recently devised copper-lined (CL) primary coil that is shown in figure 1 [7,8], the current mainly flows on the inner surface of the primary coil, which is achieved by using a copper lined inside the steel-made outer coil. Thereby, the feed-gap effect is considerably reduced and the inductive coupling between the primary coil and the liner is efficiently enhanced, which significantly raises the transfer efficiency of the electromagnetic energy into the liner. Using this coil, we recently reported the generation of magnetic fields of 730 T, which are the highest fields achieved thus far in an indoor setting [8]. In addition, various measurements for solid-state physics applications [9-11] have been performed for field above 500 T. In such an extremely high magnetic field, precise calibrations of a pick-up coil, conventionally used to detect a magnetic field, are of particular importance, because signals detected from a material are discussed based on the plots in which the abscissa is the intensity of the magnetic field. The reliability of the absolute values of the magnetic field has been investigated by means of the Faraday rotation of glasses compared with the signals from the pick-up coils and discussed in detail [12].

There is a high demand for precisely controllable EMFC techniques, which are indispensable for solid-state physics experiments performed using ultra-high magnetic fields as described above; issues of particular interest include reproducibility of a temporal profile of magnetic fields, estimation of the final size of an inner radius of the imploding liner for peak magnetic fields, and more. In this respect, detailed knowledge of dynamical magnetic flux compression processes is required for EMFC experiments. Computer simulations form a powerful tool for understanding the dynamics of the liner's imploding motion [13,14]. The simulation results, when compared carefully with the experimental data, can be useful for the understanding of not only the controllability of the magnetic field but also the progress of spatial homogeneity of the magnetic field. The latter is particularly important for solid-state physics measurements.

However, the simulations that are reported in references [13,14] have been limited to a qualitative explanation of the experimental results. The simulations have predicted a much higher peak magnetic field than what has been obtained in real experiments. The major cause of the discrepancy arose from the unrealistic assumption of a current in a primary coil neglecting the skin depth effect. Furthermore, the primary coils that were employed in those studies were rather primitive, and yielded unexpected electrical decoupling between the primary coil and the liner, which caused unfavourable effects of the feed-gap [15]. However, our newly designed CL coil could be well formulated as a simplified model to be used in simulations, which allowed more accurate and realistic prediction of our experimental results.

In this article, we discuss the dynamical process of flux compression in the EMFC experiments, based on the close comparison of simulations and experiments. We measured the time-dependent spatial distribution of the magnetic field. We observed the temporal development of a characteristic spatial distribution, and compared these results with the simulation. By using the simulations and a shadowgraph of high-speed framing camera, a more precise match of the imploding liner's motion to the real case could be identified.

**2. Experimental Methods and Simulation**

Details regarding the setup of the EMFC experiment are described in Ref. 8. Figure 1 illustrates the schematic view of the EMFC magnet that was used in this study. In the EMFC technique, the liner's implosion is caused by the electromagnetic repulsive force produced between a large current in the primary coil injected from the condenser bank and the current that is induced electromagnetically in the liner (bold arrows). The liner was set in a vacuum chamber that consisted of a bakelite (phenol resin) cylinder with transparent acryl plates on both sides, serving as its lids. We wound several pickup coils around a 1 mm diameter G10 rod, as a probe for magnetic field, to obtain the magnetic field distribution along the direction normal to that of coil's radius (defined here as the $z$-axis). For convenience of discussion, let $z = 0$ at the liner's centre. The G10 rod with pickup coils was inserted at the centre position of the imploding liner, and was fixed to one side of acryl plates. The error in the magnetic field of each pickup coil due to the fabrication processes was typically 3%, and a detailed description of the magnetic field calibration method can be found in Ref. 12.

Figure 2(a) shows a typical magnetic field curve $B(t)$ (hereafter termed the "$B$-$t$ curve") generated by the EMFC technique. The turn-around point (TA point) is defined as a point at which the $B$-$t$ curve at $z = 0$ takes the maximum value. The inset illustrates the position of pickup coils with respect to the positions of the cross section of the primary coil and the liner. In this article, we mainly report the experiment that was performed by using a 45 mm wide ($L_{pri}$) primary coil. In Sec. 3.5, the results are compared with the results that were obtained by using a primary coil with $L_{pri}$ = 90 mm. The width and thickness of the liner ($L_{lin}$, $w_{lin}$) were set to be 50 mm and 1.5 mm. The seed (initial) magnetic field to be compressed was set at 3.8 T, and 4 MJ of energy was used as an energy source stored in condenser banks, resulting in the injection of a primary current having a peak of 4.2 MA into the coil.

The shadowgraph of the imploding liner was taken by using a high-speed framing camera (IMACON 468, John Hadland). The marking-off coaxial circles with radii of 10, 20, and 30 mm were drawn on the acryl plates on both sides of the primary coil, and were used as a reference for an accurate calibration of the imploding liner's dimensions. This is schematically illustrated in figure 1 by black coaxial circles, and the inset of Figure 1 shows the enlarged view from the $z$-direction. The marking-off circles on both sides of the primary coil were also used to adjust the alignment of an optical path from a flush lump to the framing camera, in order to avoid distortion of the image due to the optical misalignment. A typical shadow image during the liner's implosion is shown in Figure 2(b) for different times. A highly symmetrical implosion of the liner was observed in all frames, indicating that the feed-gap effect is almost negligible.

To compare with the experiments, we performed the mesh modelling simulations that were based on the work of Miura and Nakao [14] (hereafter termed the MN model). The cross-section of the primary coil and the liner was divided into a fine mesh. The values of the mutual and self inductance between the mesh cells were calculated, following which the mesh current and the deformation of the liner were calculated during each time step. The procedure was described in details in Ref. 14, where the liner's thickness was assumed to be kept uniform along the $z$-direction during the implosion. This assumption is equivalent to the assumption of constant magnetic pressure all along the $z$-direction at a surface of the liner. However, this model is far from describing the real situation and is somewhat over-simplified. The magnetic pressure should strongly depend on the $z$-coordinate, because the current tends to concentrate at the objects' edges. Therefore, we revised the MN model by independently calculating the equation of motion for the liner mesh with its respective $z$-coordinate, which allows obtaining a non-uniform deformation of the liner (this revised model is hereafter referred to as the NUD model). A similar model was devised and studied by a research group from the Loughborough University [16,17]. In advance, we measured the spatial distribution of the initial magnetic field along the $z$-axis, the values of which were taken into account in our calculations.

In the calculation, $L_{lin}$ is kept constant during the implosion, and both the compressibility of the liner and the diffusion of magnetic flux into the liner are neglected. Due to the assumption of incompressibility of the liner, the volume of the liner was set to be constant. Since the lined copper plate with 2 mm thick in the CL coil bears the major part of the primary current [8], a static 2 mm thick primary coil is assumed in both model calculations. The residual impedance of a circuit was assumed to be constant for primary currents reaching up to 4 MA. Based on the experiment with a short circuit end load, the residual resistance and inductance were estimated to be 1.113 mΩ and 43.46 nH, respectively, in the case when 8 units of the condenser bank (5.0 mF) were used. The temperature dependence of the resistivity in copper was taken from Ref. 18. The values of the remaining simulation parameters were identical to the values of their corresponding experimental parameters.

## 3. Results and Discussion

*3.1. Distribution of the magnetic field along the z-axis*

We investigated in details the spatial distribution of magnetic fields inside of the liner along the z-axis. Figure 3(a) shows the *B-t* curves that were obtained from eight pickup coils whose position in the liner is shown in the inset. The pickup coil labelled "0" was set at the centre of the liner. All pickup coils succeeded in measuring the fields up to their peak intensity values without any failures on the way of elevating of the magnetic field. The respective peak field showed monotonous decrease from the centre to the edge, along the *z*-axis. In addition, the time at which the field reached its peak intensity increased with increasing *z*. It should be noted that the *B-t* curve at $z = 0$ mm attains its highest value along the *z*-axis only after 39.5 μs. This type of temporal evolution of the field's spatial distribution was reproducible in other experiments under the same conditions.

Figure 3(b) displays the temporal evolution of the magnetic field's profile in the *z*-direction that was obtained from Figure 3(a). To the best of our knowledge, this type of data has not been reported before for the EMFC or the chemical explosive flux compression experiments. Before the time of 39.7 μs, the experimental data $B(z)$ (open circles) are reproduced by fitting a curve (a dashed curve), which consists of a sum of two Gaussian curves. In figure 3(b), we used a symmetric fitting curve centred at $z = 0$ mm. A good symmetry of the magnetic field intensity centred at $z = 0$ mm was repeatedly confirmed by the *B-t* curves at $z = \pm 10$ mm (for example, points B and D in figure 2(a)), whose difference in intensity was evaluated to be 10% at its maximum. At first, the peaks of individual Gaussian curves are located at $z = \pm 10$ mm, forming a hump structure, after which they approach the point $z = 0$ mm as the time elapses, and finally merge to form a single Gaussian peak (a solid curve) at the TA point (40.1 μs).

When attempting to apply the EMFC method to the measurements of sample materials, it is very important to estimate the homogeneity of a magnetic field in a given volume of a sample. Field homogeneity, Δ*B*, was estimated from the fitting curves for $B(z)$ in figure 3(b), and the results are summarized in figure 4 by using the normalization, $\Delta B/B_{z=0}$, at $z = \pm 1$ mm and $\pm 2$ mm. The normalized value $\Delta B/B_{z=0}$ is less than 0.5% for *z* positions ranging between -1 mm and

+1 mm, where most samples can be accommodated for any kind of measurements. However, when the positions are $z = \pm 2$ mm, the field homogeneity degrades suddenly up to almost 2%.

*3.2. Simulation of the flux compression*

The two-hump structure of $B(z)$ suggests that there is a non-uniform deformation of a liner or a non-uniform distribution of liner's current, which can be compared with the results obtained from the computer simulation of an imploding liner. Figures 5(a) and 5(b) present the simulation results obtained for a temporal evolution of a cross-section of an imploding liner by using the MN model and by using the NUD model, respectively. The liner's speed is faster in the NUD model as compared with the MN model. In the MN model, the liner retains the rectangular form of its cross-section to the very end of the implosion. On the other hand, in the NUD model, the liner's cross-section develops as a spool-like form. The liner starts with its edge ($z \sim 20$ mm) being slightly deformed, and the central section of the liner implodes preferentially. At the TA point (37.3 µs), the liner retains almost a rectangular form. A little protrusion can be noticed on the inner surface of the liner at $z = 0$, indicating that the inner diameter of the liner takes on its minimum value at $z = 0$ along the $z$-axis at the TA point. Due to the non-uniform deformation of the liner, as shown in figure 5(b), the thickness obtained from the shadowgraph of the liner (defined by $w_1 \equiv r_{max} - r_{min}$) differs from the actual thickness near $z = 0$ (defined by $w_0 \equiv r_o - r_i$). Here, $r_{max}$ and $r_{min}$ are the outer and inner radius taken from the shadowgraph of the framing camera (figure 2(b)), while $r_o$ and $r_i$ are those at the centre $z = 0$. The difference between $w_0(t)$ and $w_1(t)$ will be discussed in Sec. 3.3.

The colour plot in figure 5 shows the distribution of currents in an imploding liner, $I_s$. At first, $I_s$ takes negative values in the majority of points on the mesh, reflecting the fact that $I_s$ is induced by the primary current. Then, $I_s$ changes its sign, mainly inside the liner, to confront with the compressing magnetic flux. This temporal evolution of $I_s$ in the NUD model is qualitatively similar to the one that is observed in the MN model [14]. However, the faster implosion of the liner in the NUD model leads to the fast temporal evolution of the positive $I_s$ at the inner surface of the liner (already observed at 25 µs), which yields the faster magnetic field confinement, as compared with the MN model. Furthermore, in the case of the NUD model the current $I_s$ is more efficiently concentrated on the inner surface of the liner, as compared with the MN model. In the NUD model, 90% of the total $I_s$ is bore by the mesh at the 1$^{st}$ layer with a typical 1 mm thickness of the liner, whereas this value reaches only 87.5% for the MN model.

In figure 6(a), the *B-t* curves are compared with the simulated curves that were mentioned above (bold curve: experiment, thin curves: the NUD model and the MN model). The *B-t* curve of the NUD model reaches its peak intensity much closer in time to the experimental peak than the MN model. Yet, in both models, the intensities of the peak field are much better reproduced than those that were reported in Ref. 14. These results indicate that the CL primary coil is ideally designed to yield a good match with model calculations. Figures 6(b) and 6(c) illustrate the temporal evolution of $B(z)$, produced from the simulations by (b) the NUD model and (c) the MN model, whose respective time points are indicated by open circles on the $B$-$t$ curves in figure 6(a). The field $B(z)$ of the NUD model strongly depends on the $z$-coordinate, compared with the $B(z)$ of the MN model, and reflects the liner's shape deformation during the implosion, as shown in figure 5(b).

The $B(z)$ curve strongly depends on the liner's shape during implosion. Next, we focus on the liner's deformation with respect to the relation between $L_{pri}$ and $L_{lin}$. Due to the non-uniform spatial distribution of the Maxwell stress that is exerted on the liner, reflecting high current concentration on the liner's edges, a drastic difference in a deformation of the imploding liner is anticipated, depending on the size of $L_{lin}$ relative to $L_{pri}$. Therefore, we simulated two distinctive cases: (a) $L_{lin}$ = 35 mm ($L_{lin} < L_{pri}$), and (b) $L_{lin}$ = 55 mm ($L_{lin} > L_{pri}$), as shown in figure 7. The colour map shows the distribution of $I_s$. For the primary current $I_p$, the value at time 10 μs is shown on the right-hand side of figure. The concentration of $I_p$ at the edges is weaker in the case of $L_{lin} < L_{pri}$, as compared with the case of $L_{lin} > L_{pri}$. Therefore, it can be inferred that for $L_{lin} < L_{pri}$ the Maxwell stress is less effective at the liner's edge, resulting in the barrel-like deformation of a liner. On the other hand, for $L_{lin} > L_{pri}$ a more effective influence of the Maxwell stress on the edges of the liner results in a spool-like deformation and relatively faster implosion. In addition, there was a prominent efficient sign reversal of $I_s$ only at the inner surface of the liner. Therefore, we stress that the compression of a magnetic field strongly depends on the liner's deformation during implosion. As is noted in figure 7(c), a two-hump structure appears in $B(z)$, which indicates that $B(t, z = 0)$ does not take the maximum value along the $z$-direction before reaching the TA point (38.1 μs). This result is similar to the experimental result for times of up to 39.5 μs in figure 3(b), whereas for $L_{lin} > L_{pri}$ (figure 7(d)) $B(z)$ significantly differs from the corresponding experimental result. In figure 7(c), the hump structures are located almost near the edges ($z \sim 12.5$ mm). Provided that $L_{lin}$ is kept constant, the peak position of a hump remains the same at all times, as shown by figure 7(c). In order to reproduce the observed humps with the gradual approach of two peaks as can be seen in figure 3(b), the parameter $L_{lin}$ cannot be kept constant, but should be decreased with the liner's implosion before reaching the TA point.

*3.3. Comparison with the liner framing shadowgraph*

The liner's shape that is captured by the framing camera provides important information regarding the liner's implosion process. We estimated the quantities $r_{max}$ and $r_{min}$ from the shadowgraph, and compared with those obtained from the simulation of the NUD model in figure 8 (labelled as $r_{max,e}$ and $r_{min,e}$). The experimental data in figure 8 were obtained from the experiment performed with the same conditions as used in the case of the results shown in figure 3. Figure 8(a) shows the liner's shadowgraph thickness $w_{1,e} \equiv r_{max,e} - r_{min,e}$ (triangles) together with the simulated quantities $w_1$ and $w_0$ using the NUD model. The experimental data and the quantity $w_1$ are in a very good agreement up to the time of 22 μs, showing an almost twofold increase in the thickness $w_0$. This result strongly supports the non-uniform deformation of the liner that was discussed above.

Furthermore, as shown in figure 8(b), there exists an important observable indication of the liner's implosion near the TA point. We noted that $r_{max,e}$ (open squares) continued to decrease further by about 1 μs, even after reaching the TA point (40.7 μs, as indicated by an arrow). Based on the NUD model simulations, we found that the observed decrease in $r_{max,e}$ after the TA point arises due to the delayed implosion of the liner's edge. Figure 8(b) shows that $r_i$ (closed downward triangles) coincides with $r_{min}$ (closed circles) up to the TA point, owing to a little protrusion of an inner surface of the liner at $z = 0$, as has already been discussed for figure 5(b). On the other hand, due to the delayed implosion of the liner's edge, $r_{max}$ (closed squares) is larger than $r_o$ (closed upward triangles). Therefore, $r_{max}$ continues to decrease further by about 1 μs after the TA point, where $r_{min}$ attains its minimum value. This behaviour also constitutes an additional evidence of the non-uniform deformation of the liner. Very close to the TA point, reliable evaluation of $r_{min,e}$ (open circles) becomes difficult due to the resolution of a framing camera and the explosive halation at the inner surface of the liner (for example, see 6[th] frame in figure 2(b)).

The delayed implosion of liner's edge after the TA point also affects the $B$-$t$ curves with different $z$-coordinates. By comparing the results of the NUD simulation (figure 7), we found that the $B$-$t$ curve has a different dependence on the $z$-coordinate for the two cases. As noted in figure 9(a) for $L_{lin} < L_{pri}$, the peak of $B(t, |z| > 0)$ is reached at an earlier time compared with that of $B(t, z = 0)$ as shown in figure 9(a), due to the barrel-like implosion that accompanies the two-hump structure in $B(z)$. However, completely opposite results are obtained for $L_{lin} > L_{pri}$. As is shown in figure 9(b), the peak of $B(t, |z| > 0)$ is always delayed relative to that of $B(t, z = 0)$. This type of delay was observed in figure 3(a) after the TA point. By comparing these two different cases, we infer that the liner's shape during the implosion changes from the spool-like type to the barrel-like type with decreasing $L_{lin}$; then, $L_{lin}$ increases again near the TA point.

*3.4. A scenario of the liner's imploding process*

The discussion above led us to consider the following scenario to describe the liner's imploding process. The scenario is summarized in figure 10 with the *B-t* curve. The inset shows the schematic illustration of the temporal evolution of the liner's cross-section. Initially, the liner favours a spool-like shape (phase I). Then, the liner gradually shrinks also along the *z*-axis and deforms to a barrel-like shape, which results in the two-hump structure in *B*(*z*) and a rapid increase in $w_{1,e}$ (phase II). Near the TA point, the liner becomes less influenced by the edge current in the primary coil, and the magnetic pressure acting on the liner's inner surface becomes strong. Therefore, the liner's barrel-like shape relaxes, accompanied by the outwards stretching deformation of the liner along the *z*-axis (phase III), where the peak of *B*(*t*, |*z*| > 0) (points B and C in figure 10) occurs later in time than the TA point. The liner's deformation along the *z*-axis during the implosion may become feasible if Maxwell stress component along the *z*-axis is taken into consideration, which is ignored in our simulations. The above scenario can explain the overall features of our experimental data.

*3.5. Comparison with the wide primary coil*

Finally, we discuss the results obtained by using a primary coil with $L_{pri}$ = 90 mm, which is twice the value that was employed above. The experimental conditions were slightly different. The seed field was 3.56 T, and the energy of 4.5 MJ was used as an energy source stored in condenser banks, resulting in the injection of a 5.7 MA peak primary current into the coil. As shown in figure 11(a), a much higher peak intensity of magnetic field is expected to be obtained from the simulations than what is obtained by using the primary coil with $L_{pri}$ = 45 mm, due to a better energy transfer efficiency associated with an enhanced barrel-type liner's implosion (inset). However, unlike in the case of $L_{pri}$ = 45 mm, we have failed to experimentally obtain the magnetic fields up to the TA point. In figure 11(b), a pickup coil at *z* ~ 10 mm (bold solid curve) is broken at the early stage of implosion (37 μs), whereas *B*(*t*, *z* ~ 10 mm) attains its peak magnetic field intensity of 426 T at 40.8 μs for $L_{pri}$ = 45 mm (pickup coil labelled "3" in figure 3(a)). For $L_{pri}$ = 90 mm, early destruction was also observed for *z* = 0 (bold dotted curve), and explosive halation in the flaming camera took place at 37.6 μs ("X" point in the figure). At around 27.6 μs, a liner's thickness of the $L_{pri}$ = 90 mm coil (calculated from open symbols, $r_{max}$ and $r_{min}$) was found to be much larger than that of the $L_{pri}$ = 45 mm coil (calculated from solid symbols, $r_{max}$ and $r_{min}$), which reflects the simulation results in the inset of figure 11(a). Therefore, in the case of the primary coil with $L_{pri}$ = 90 mm, the imploding liner might prematurely crash due to the enhanced barrel-like deformation. These results suggest the importance of a 3-dimensional deformation of a liner during the process of implosion in selecting the coil parameters for further development of the EMFC technique aiming at higher magnetic fields.

## 4. Conclusion

   The spatial distribution of the magnetic fields generated by the EMFC technique and the shadowgraph of a liner captured by a framing camera were analysed using computer simulations, focusing on the aspects of dynamical processes in an imploding liner. The simulated temporal evolution of the magnetic field intensity is in a good agreement with the experimentally obtained one. We clarified, for the first time, the close connection between the magnetic field's spatial distribution and the non-uniform implosion process of a liner, which includes the inwards compression and subsequent outwards stretching motion of the liner along the magnetic field axis. The present study led to a conclusion that a detailed study of liner's deformation in an imploding process is of particular importance for the increased controllability and the generation of higher magnetic fields than that of the present state-of-the-art record.

**References**

[1] Miura N and Herlach F 1985 *Pulsed and Ultrastrong Magnetic Fields*, *Springer Topics in Applied Physics* Vol. 57, ed F Herlach (Springer, Berlin) p 247

[2] Herlach F 1999 *Rep. Prog. Phys.* **62** 859

[3] Miura N and Herlach F 2003 *High Magnetic Fields: Science and Technology Volume 1: Magnet Technology and Experimental Techniques*, ed F Herlach and N Miura (Singapore, World Scientific) p 235, and reference therein

[4] Fowler C M, Garn W B, and Caird R S 1960 *J. Appl. Phys.* **31** 588

[5] Bykov A I, Dolotenko M I, Kolokolchikov N P, Selemir V D, and Tatsenko O M 2001 *Physica B* **294-295** 574

[6] Cnare E C 1966 *J. Appl. Phys.* **37** 3812

[7] Takeyama S, Sawabe H, and Kojima E 2010 *J. Low Temp. Phys.* **159** 328

[8] Takeyama S and Kojima E 2011 *J. Phys. D: Appl. Phys.* **44** 425003

[9] Sekitani T, Miura N, Ikeda S, Matsuda Y H, and Shiohara Y 2004 *Physica B* **346-347** 319

[10] Takeyama S, Suzuki H, Otsubo Y, Yokoi H, Murakami Y, and Maruyama S 2011 *J. Phys.: Conf. Ser.* **334** 012052

[11] Miyata A, Ueda H, Ueda Y, Sawabe H, and Takeyama S 2011 *Phys. Rev. Lett.* **107** 207203

[12] Nakamura D, Sawabe H, Matsuda Y H, and Takeyama S 2013 *Rev. Sci. Instr.* **84** 044702

[13] Miura N and Chikazumi S 1979 *Jpn. J. Appl. Phys.* **18** 553

[14] Miura N and Nakao K 1990 *Jpn. J. Appl. Phys.* **29** 1580

[15] Matsuda Y H, Herlach F, Ikeda S, and Miura N 2002 *Rev. Sci. Instr.* **73** 428

[16] Novac B M, Smith I R, and Hubbard M 2004 *IEEE Trans. Plasma Sci.* **325** 1896

[17] Novac B M and Smith I R 2006 *Jpn. J. Appl. Phys.* **45** 2807

[18] Matula R A 1979 *J. Phys. Chem. Ref. Data* **8** 1147

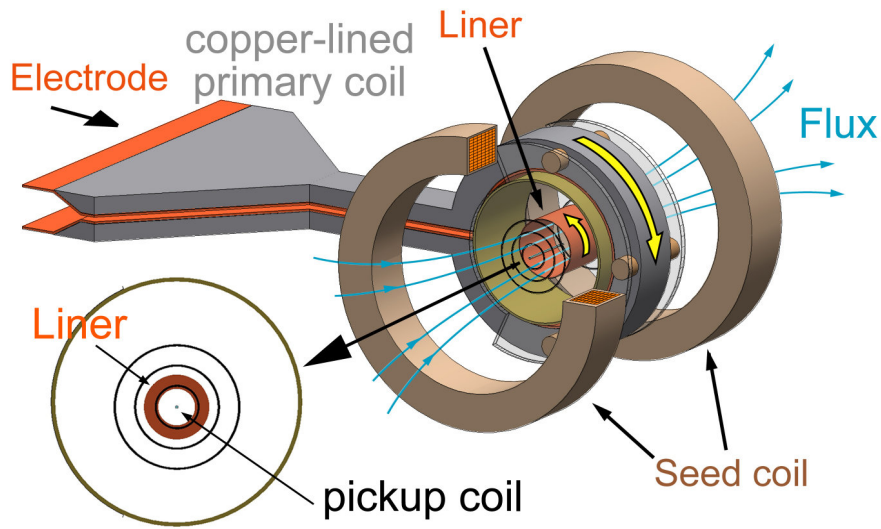

**Figure 1.** Schematic of the EMFC magnet that is composed of a copper-lined primary coil with a copper liner inside and a pair of seed field coils. The inset is the enlarged view of the cross-section from the magnetic field axis.

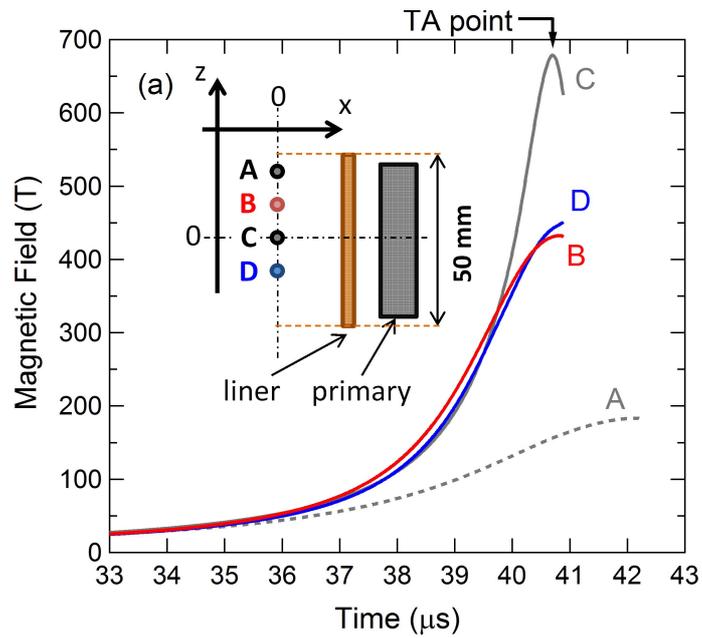

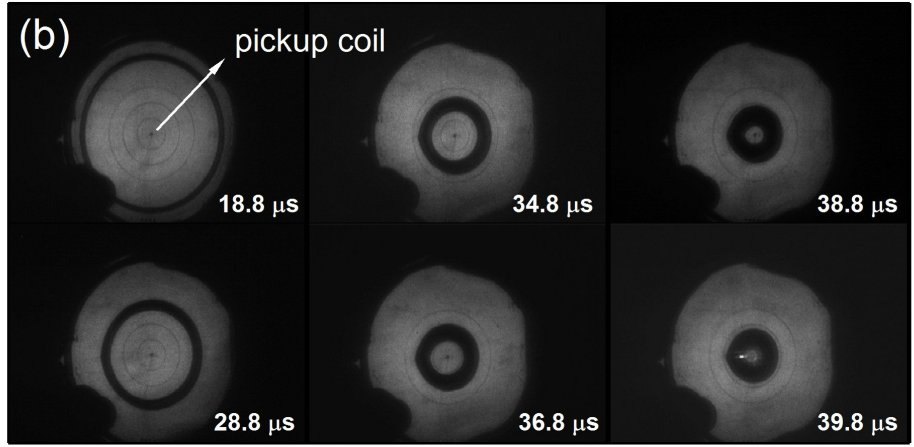

**Figure 2.** (a) Typical curves of magnetic field intensity along the *z*-axis [*z* ~ (A) 20 mm, (B) 10 mm, (C) 0 mm, and (D) -10 mm]. Inset illustrates the position of pickup coils relative to the cross-section of the primary coil and the liner. (b) A typical shadowgraph of the imploding liner, captured by a framing camera.

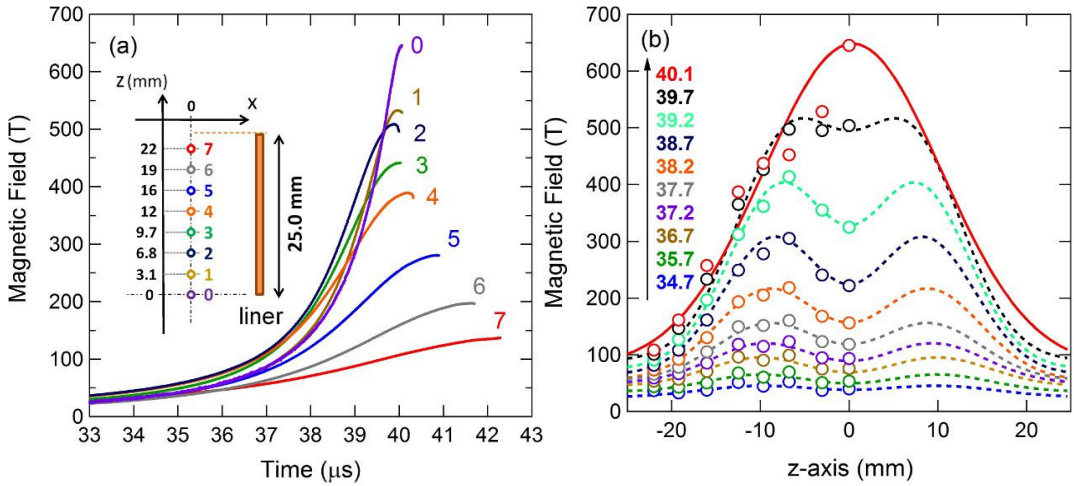

**Figure 3.** (a) *B-t* curves measured in eight pickup coils placed along the *z*-axis. Inset illustrates positions of pickup coils relative to the cross-section of half of the liner. (b) Spatial profiles of magnetic field intensity at different times (numbers in bold, time is measured in microseconds). Solid curve is the single Gaussian function, and dashed curves are the summations of 2-Gaussian functions.

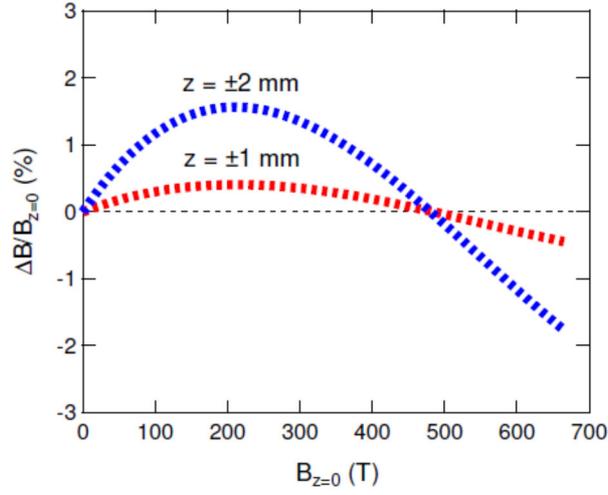

**Figure 4.** Relative deviation of a magnetic field with increasing magnetic field at $|z| = 1$ mm and 2 mm, respectively, normalized by $B(z = 0)$.

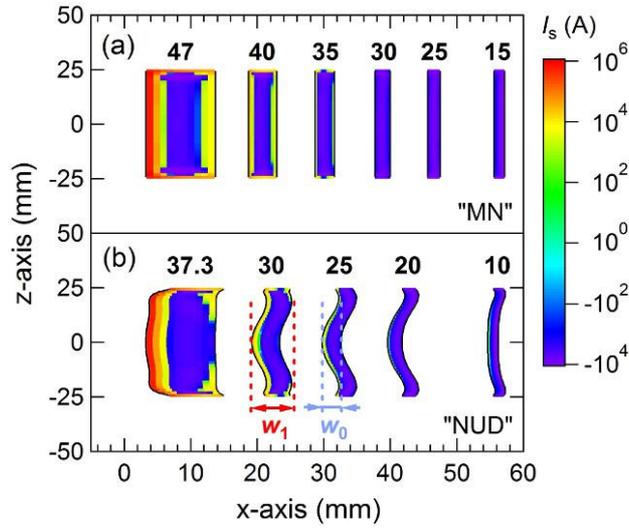

**Figure 5.** The liner's deformation calculated by (a) the MN model and (b) the NUD model at different times (numbers in bold, time is measured in microseconds). The colour plot indicates the current inside the liner, $I_s$. In (b), $w_0$ is the thickness at $z = 0$, and $w_1$ is the apparent contour that appeared in a shadowgraph.

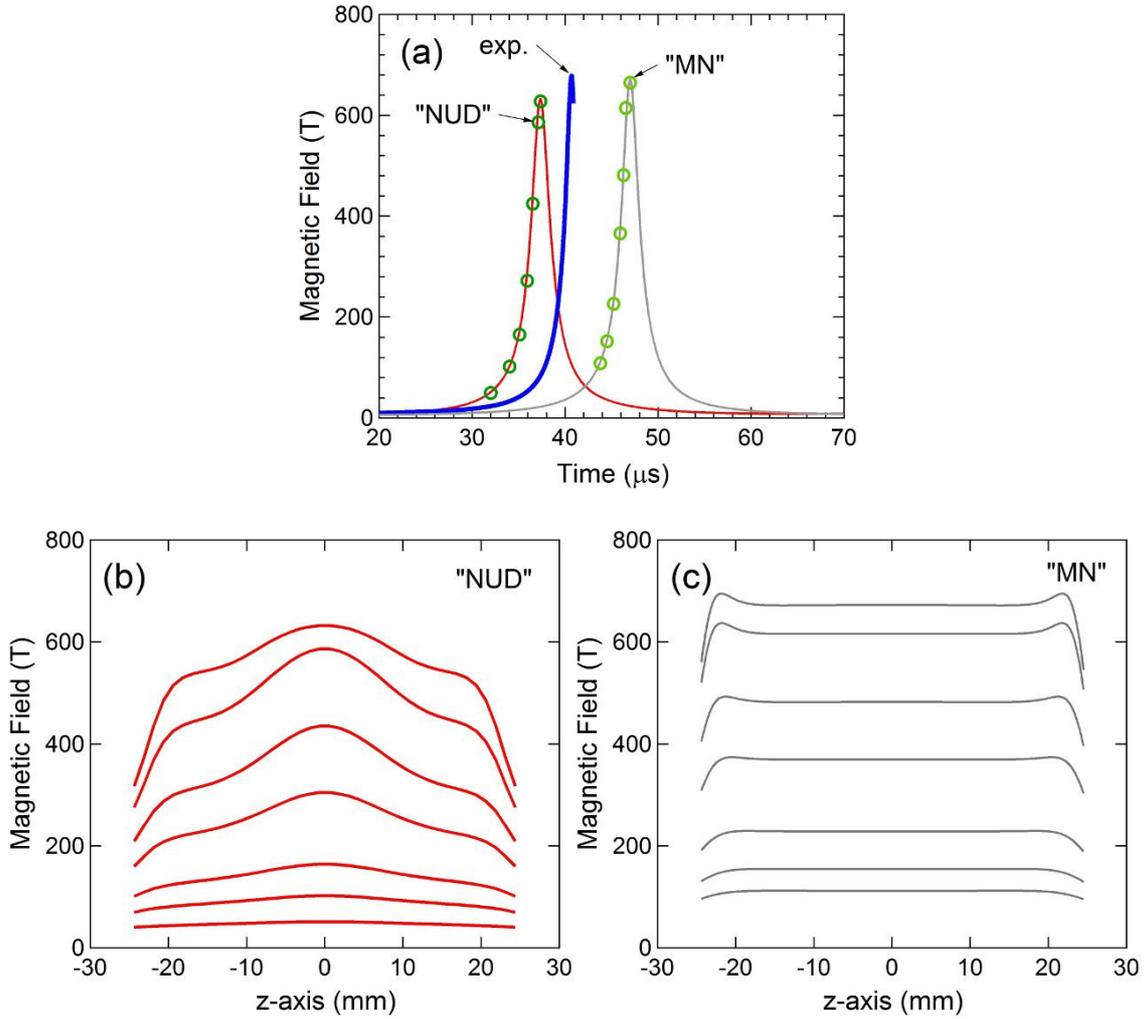

**Figure 6.** (a) Comparison of the experimental *B-t* curve (bold curve) with the curves that were obtained from simulations (thin curves: NUD model and MN model). (b),(c) The spatial dependence of magnetic field of (b) the NUD model and (c) the MN model, at different time points that are marked as circles in (a), respectively.

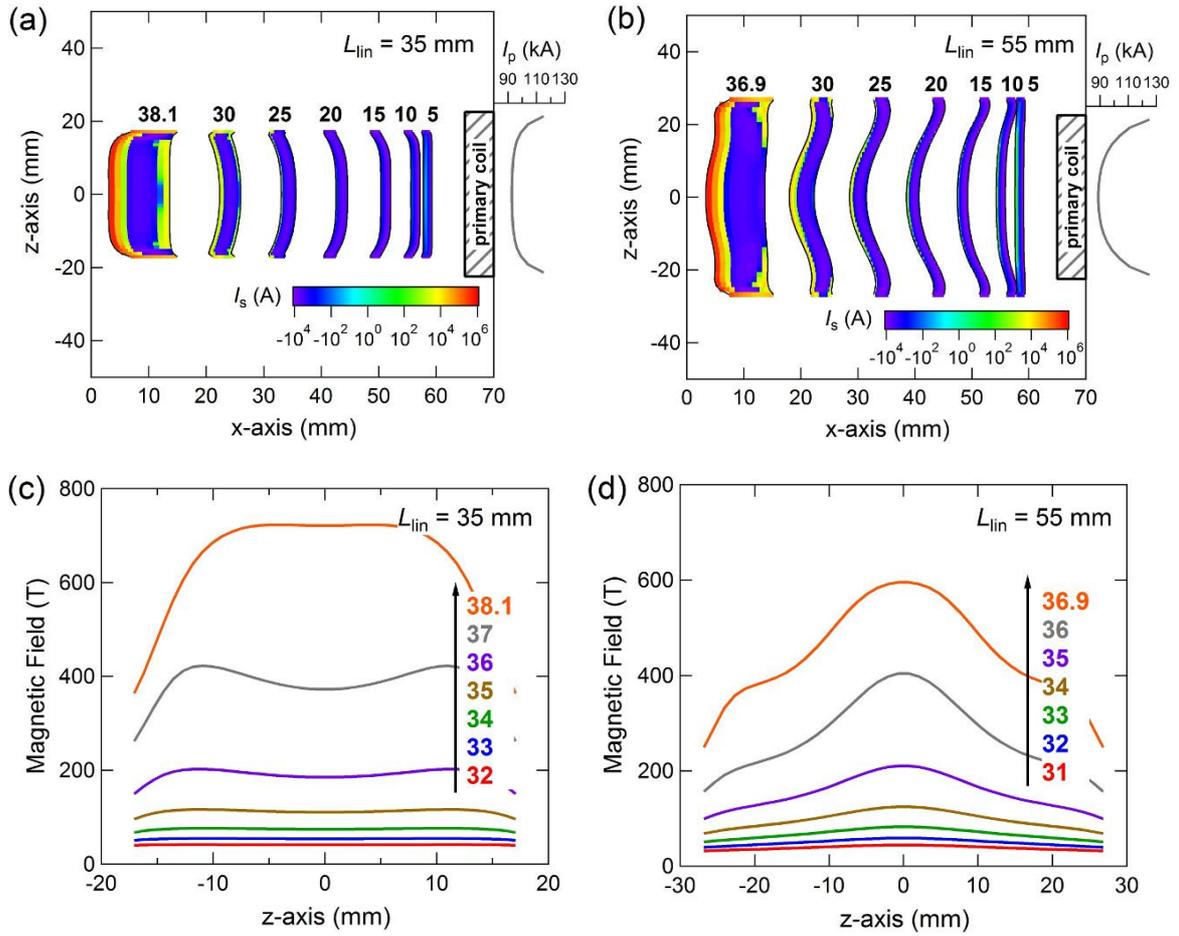

**Figure 7.** Simulation results, obtained for different liner widths $L_{lin}$. Non-uniform deformation of the liner is shown at different times (numbers in bold, time is measured in microseconds) for $L_{lin}$ = (a) 35 mm, (b) 55 mm. The colour maps show the distribution of the liner's current $I_s$. The distribution of the primary current $I_p$ is shown on the right-hand side of each figure. The spatial dependence of the magnetic field is shown for $L_{lin}$ = (c) 35 mm, (d) 55 mm, for different times (numbers in bold, time is measured in microseconds).

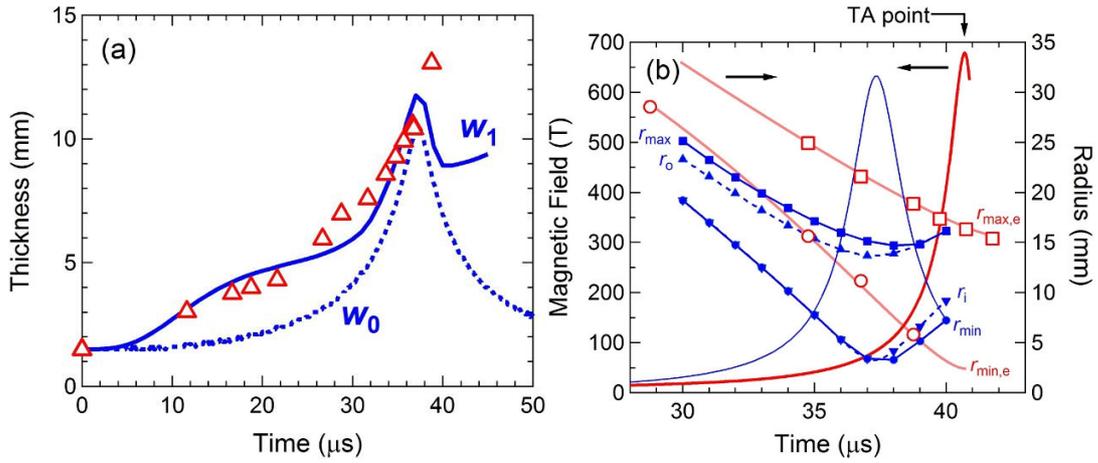

**Figure 8.** (a) The thickness $w_{1,e}$ (triangles) of the liner's shadowgraph compared with $w_0$ (dotted curve) and $w_1$ (solid curve) that were obtained from the simulation of NUD model. (b) The open symbols are for the shadowgraph data (squares: $r_{max,e}$, circles: $r_{min,e}$), with the curves serving as eye-guides. The solid symbols are for the simulated data (squares: $r_{max}$, circles: $r_{min}$, upward triangles: $r_o$, downward triangles: $r_i$). The experimental $B$-$t$ curve is also plotted by a thick solid curve, and a thin solid curve represents the simulated $B$-$t$ curve.

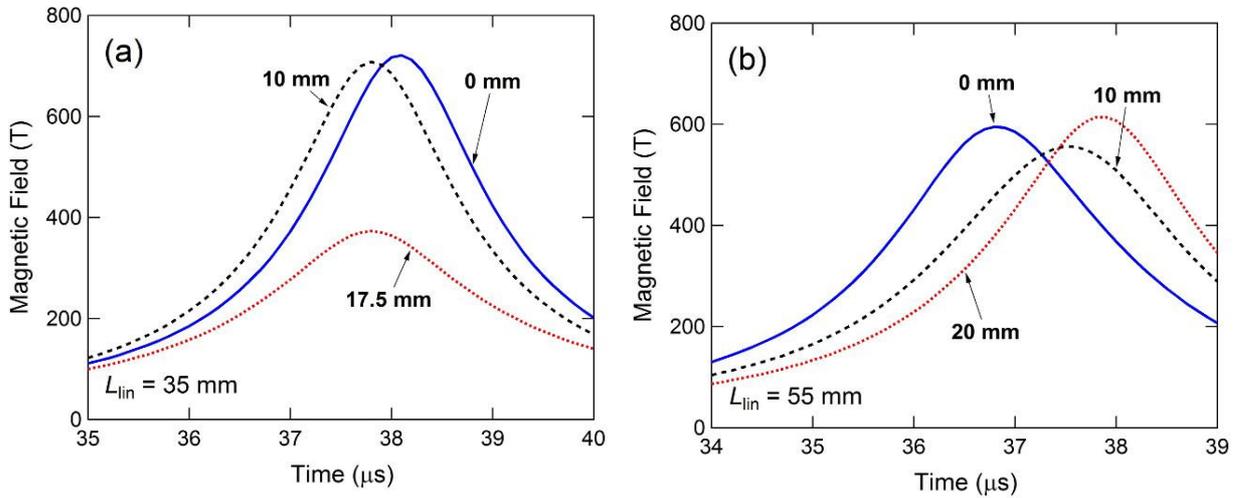

**Figure 9.** The $B$-$t$ curves for different positions on the $z$-axis, calculated by using the NUD model for: (a) $L_{lin} = 35$ mm ($L_{lin} < L_{pri}$), and (b) $L_{lin} = 55$ mm ($L_{lin} > L_{pri}$).

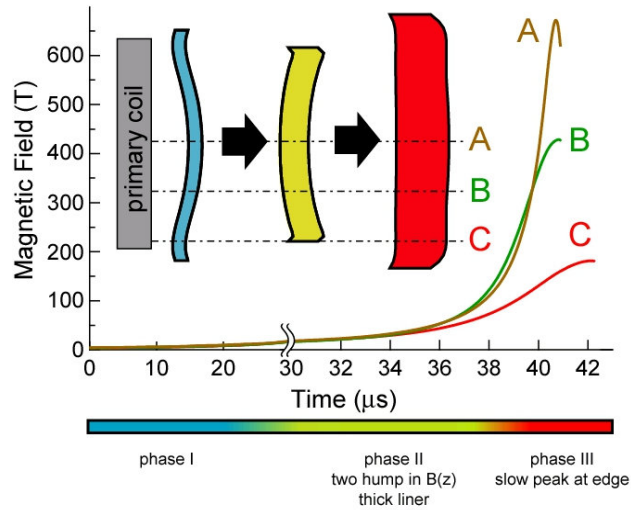

**Figure 10.** A scenario of the liner's imploding process. Inset illustrates the deformation of the liner in each phase.

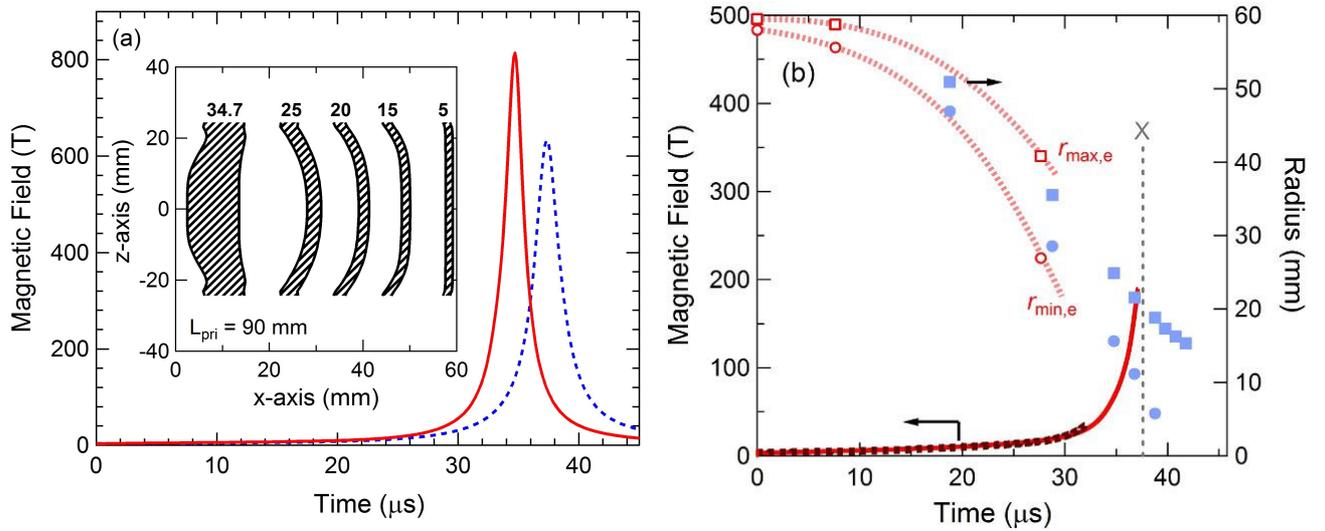

**Figure 11.** (a) Comparison of the simulation results obtained for $L_{pri} = 45$ mm (dashed curve), and 90 mm (solid curve). Inset shows the liner's deformation for $L_{pri} = 90$ mm at different times (numbers in bold, time is measured in microseconds). (b) Comparison of the experimental results of $r_{min}$ and $r_{max}$ for $L_{pri} = 45$ mm (solid symbols), and 90 mm (open symbols). At a point X, explosive halation was observed in the framing camera. The $B$-$t$ curve for $L_{pri} = 90$ mm at $z \sim 0$ mm (bold dashed curve) and 10 mm (bold solid curve) are also shown.